# Revealing common artifacts due to ferromagnetic inclusions in highly-oriented pyrolytic graphite

M. Sepioni, R. R. Nair, I-Ling Tsai, A. K. Geim, I. V. Grigorieva[*]

School of Physics and Astronomy, University of Manchester, Manchester M13 9PL, UK

*We report on an extensive investigation to figure out the origin of room-temperature ferromagnetism that is commonly observed by SQUID magnetometry in highly-oriented pyrolytic graphite (HOPG). Electron backscattering and X-ray microanalysis revealed the presence of micron-size magnetic clusters (predominantly Fe) that are rare and would be difficult to detect without careful search in a scanning electron microscope in the backscattering mode. The clusters pin to crystal boundaries and their quantities match the amplitude of typical ferromagnetic signals. No ferromagnetic response is detected in samples where we could not find such magnetic inclusions. Our experiments show that the frequently reported ferromagnetism in pristine HOPG is most likely to originate from contamination with Fe-rich inclusions introduced presumably during crystal growth.*

There have been many reports of weak room-temperature ferromagnetic signals observed in pristine HOPG [1-8]. However, the measured values of magnetization are very small, $\sim 10^{-3}$ emu/g, i.e., 5 orders of magnitude less than the saturation magnetization of Fe, and the whole subject remains controversial, especially concerning the role of possible contamination, as well as the mechanism responsible for the strong interaction required to lead to ferromagnetic ordering at room temperature.

Trying to clarify the situation, we have carried out extensive studies of magnetic behaviour of HOPG crystals obtained from different manufacturers (ZYA-, ZYB-, and ZYH-grade from NT-MDT and SPI-2 and SPI-3 from SPI Supplies). These crystals are commonly used for studies of magnetism in graphite; e.g., ZYA-grade crystals were used in refs. [1,3-7] and ZYH-grade in ref. [2]. We have also observed weak ferromagnetism, independent of temperature between 300K and 2K and similar in value to the one reported previously for pristine (non-irradiated) HOPG. Below, we show that the observed ferromagnetism in ZYA-, ZYB-, and ZYH-grade crystals is due to micron-sized magnetic inclusions (containing mostly Fe), which can easily be visualized by scanning electron microscopy (SEM) in the backscattering mode. Without the intentional use of this technique, the inclusions are easy to overlook. No such inclusions were found in SPI crystals and, accordingly, in our experiments these crystals were purely diamagnetic at all temperatures (no ferromagnetic signals at a level of $10^{-5}$ emu/g).

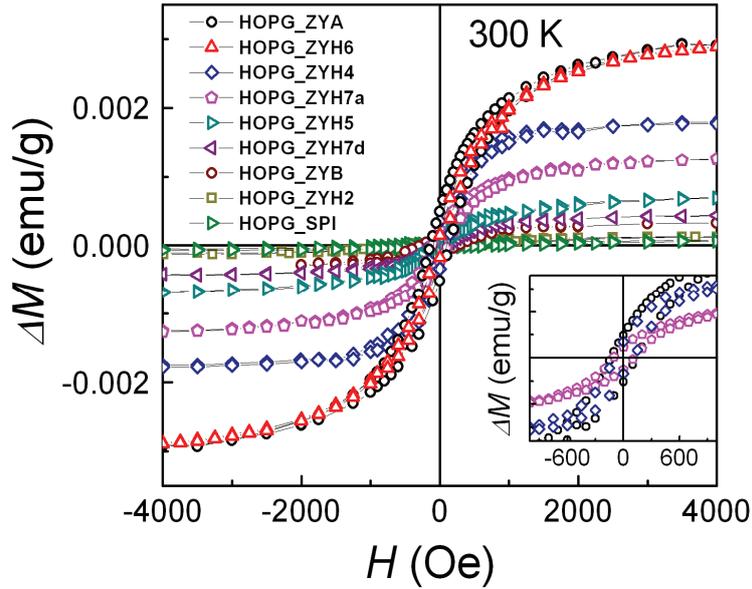

**Figure 1**. Ferromagnetic response in different HOPG crystals. Magnetic moment $\Delta M$ vs applied field $H$ after subtraction of the linear diamagnetic background. The inset shows a low-field zoom of three curves from the main panel where the remnant $\Delta M$ and coercive force are seen clearly.

Ten HOPG crystals of different grades (ZYA, ZYB, ZYH and SPI) were studied using SQUID magnetometry (Quantum Design MPMS XL7), XRFS, SEM and chemical microanalysis by means of energy-dispersive X-ray spectroscopy (EDX). For all ZYA, ZYB and ZYH crystals, magnetic moment vs field curves, $M(H)$, showed characteristic ferromagnetic hysteresis in fields below 2000 Oe, which was temperature independent between 2K and room $T$, implying a Curie temperature significantly above 300K. The saturation magnetisation $M_S$ varied from sample to sample by more than 10 times, from $1.2 \cdot 10^{-4}$ emu/g to $3 \cdot 10^{-3}$ emu/g – see Fig. 1. This is despite the fact that XRFS did not detect magnetic impurities in any of our HOPG crystals (with a detection limit better than a few ppm). This result is similar to the findings of other groups (e.g. refs. [3-5]). Figure 1 also shows an $M(H)$ curve for one of the SPI crystals, where no ferromagnetism could be detected.

The seemingly random values of ferromagnetic signal in nominally identical crystals could be an indication that the observed ferromagnetism is related to structural features of HOPG, such as grain boundaries, as suggested in ref. [2]. However, we did not find any correlation between the size of the crystallites making up HOPG crystals and/or their misalignment and the observed $M_S$. For example, the largest $M_S$ as well as the largest coercive force, $M_C$, were found for one of the ZYA crystals, which have the smallest mosaic spread (0.4-0.7°), and for a ZYH crystal with the largest mosaic spread (3-5°). Furthermore, crystallite sizes were rather similar for all ZYA, ZYB and ZYH crystals (see Fig. 2) while $M_S$ varied by a factor of 10 (Fig. 1). SPI crystals had similar mosaic spreads to ZYB and ZYH (0.8°±0.2° for SPI-2 and 3.5°±1.5° for SPI-3) and tended to have a larger proportion of smaller grains (Fig. 2c), yet did not show any ferromagnetism at all (Fig. 1).

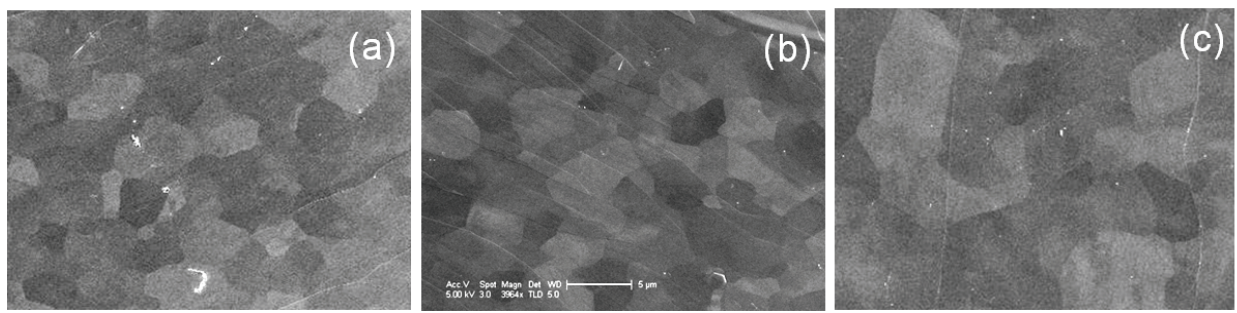

**Figure 2**. Typical, same-scale, SEM images of crystallites in different HOPG samples: (a) ZYH; (b) ZYA; (c) SPI. Typical crystallite sizes in ZYH, ZYB and ZYA are 2 to 5 μm; in SPI crystallites vary from 0.5 to 15 μm. The scale bar corresponds to 5μm.

To investigate whether the observed ferromagnetism is homogeneous within the same commercially available 1cm×1cm×0.2cm HOPG crystal, we measured magnetisation of four samples cut out from the same ZYH crystal as shown in the inset of Fig. 3. To exclude possible contamination of the samples due to exposure to ambient conditions, both exposed surfaces were cleaved and the edges cut off just before the measurements. Surprisingly, we found significant variations of the ferromagnetic signal between these four nominally identical samples – see Fig. 3. This indicates that the observed ferromagnetism is not related to structural or other intrinsic characteristics of HOPG crystals, as these are the same for a given crystal. Therefore, it seems reasonable to associate the magnetic response with external factors, such as, for example, the presence of small inclusions of another material.

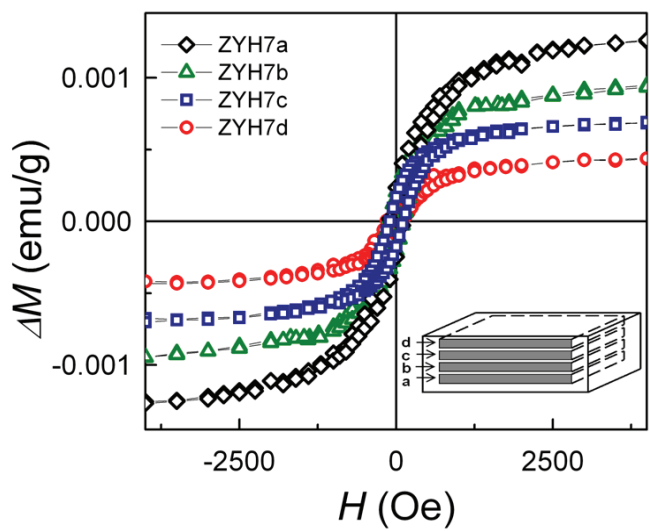

**Figure 3**. Ferromagnetic hysteresis in four samples cut from the same ZYH HOPG crystal. The inset shows schematically positions of the samples in the original crystal.

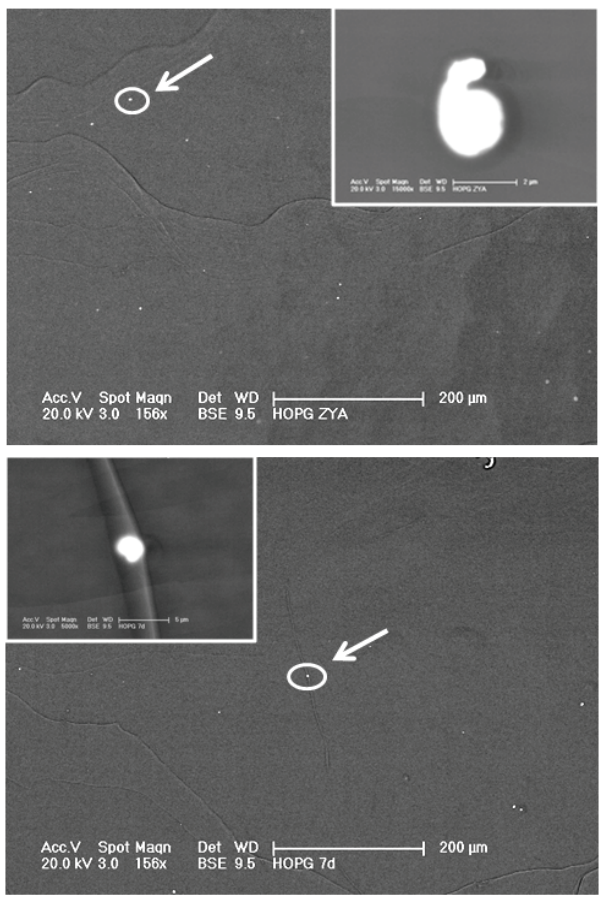 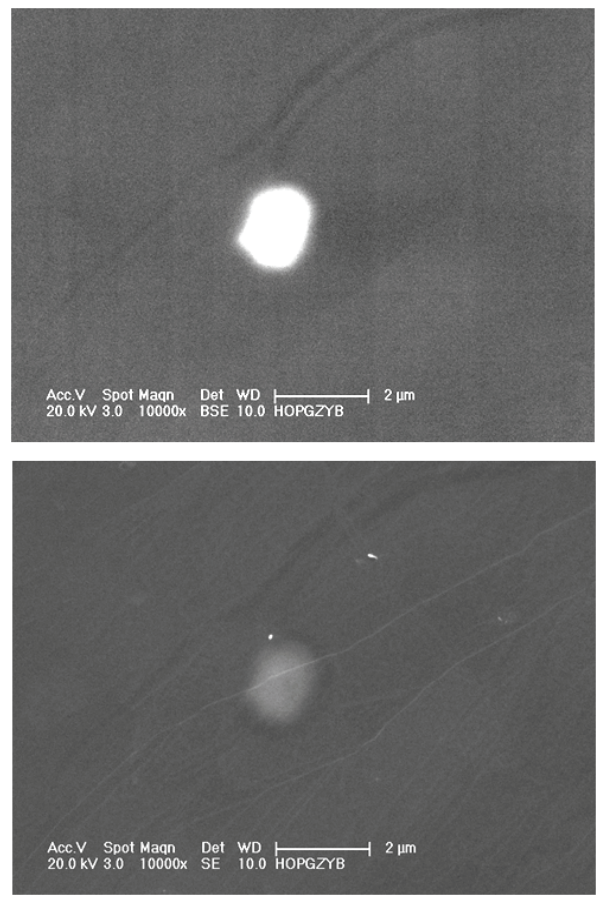

**Figure 4.** SEM images of ZYA (top) and ZYH (bottom) samples in back-scattering mode. Small white particles are clearly visible in both images, with typical separations between the particles of 100 μm for ZYA and 240 μm for ZYH. Insets show zoomed-up images of the particles indicated by arrows; both particles are ≈2μm in diameter.

**Figure 5.** SEM images of the same particle found in a ZYB sample taken in backscattering (top) and secondary electron (bottom) modes. Surface features are clearly visible in the SE image while BS is mostly sensitive to chemical composition. The contrast around the particle in the SE mode is presumably due to a raised surface in this place.

To check this hypothesis, we examined samples of different HOPG grades using backscattering mode in a scanning electron microscope (SEM). Due to their sensitivity to the atomic number [9], backscattered electrons can provide a strong contrast allowing to detect particles made of heavy elements inside a light matrix (graphite in our case). This experiment revealed that all ZYA, ZYB, and ZYH crystals contained sparsely distributed micron-sized particles of a large atomic number, with typical in-plane separations of 100 to 200 μm – see Fig. 4. Comparison of SEM images in backscattering and secondary electron modes (BS and SE, respectively) revealed that in most cases the particles were buried under the surface of the sample and, therefore, were not visible in the most commonly used secondary electron mode. This is illustrated in Fig. 5 which shows the same area of a ZYB sample in the SE and BS modes. The difference between the two images is due to different energies and penetration depths for secondary and backscattered electrons: The energy of BS electrons is close to the primary energy, i.e. ~20 keV in our case, and they probe up to 1μm thick layer at the surface [9] while secondary electrons have characteristic energies of the order of 50 eV and come from a thin surface layer of a nm thickness [10]. Importantly, no such inclusions could be detected in SPI samples that, as discussed, did not show any ferromagnetic response. The difference between ZY and SPI grades is presumably due to different manufacturing procedures used by different suppliers. Our attempts through NT-MDT to find out the exact procedures used for production of ZY grades of HOPG were unsuccessful.

To analyse the chemical composition of the detected particles we employed *in situ* energy-dispersive X-ray spectroscopy (EDX) that allows local chemical analysis within a few $\mu m^3$ volume. Figure 6 shows a typical EDX spectrum collected from a small volume (so-called interaction volume) around a 2.5 $\mu m$ diameter particle in a ZYA sample. This particular spectrum corresponds to the presence of 8.6 wt% (2.1 at%) Fe, 2.3 wt% (0.65%) Ti, 1.8 wt% V (0.47 at%) and <0.5 wt% Ni, Cr and Co, as well as some oxygen, which appear on top of 86 wt% (96.5 at%) of carbon. The latter contribution is attributed to the surrounding graphite within the interaction volume. To determine the actual composition of the inclusion, we needed to take into account that the above elemental analysis applies to the whole interaction volume, where the primary electrons penetrate into the sample. Given that 96% of the interaction volume is made up by carbon, the electron range $R$ and, accordingly, the interaction volume can be estimated to a good approximation using the Kanaya-Okayama formula [11]:

$$R = \frac{0.0276 \cdot A \cdot E^{1.67}}{Z^{0.89} \cdot \rho} \approx 4.5 \mu m,$$

where $A$=12 g/mole is the atomic weight of carbon, $E$=20 keV the beam energy, $Z$=6 the atomic number and $\rho$=2.25 g/cm$^3$ the density of graphite. Using the calculated value of $R$, the weight percentages for different elements from the spectrum and their known densities, it is straightforward to show that the volume occupied by the detected amount of Fe and Ti is in excellent agreement with the dimensions of the particle in Fig. 6, i.e., the particle is made up predominantly of these two elements. The presence of oxygen indicates that Fe and Ti are likely to be in an oxidised state, i.e. the particle is either magnetite or possibly titanomagnetite, both of which are ferrimagnetic, with saturation magnetisation $M_S \approx$ 75-90emu/g [12].

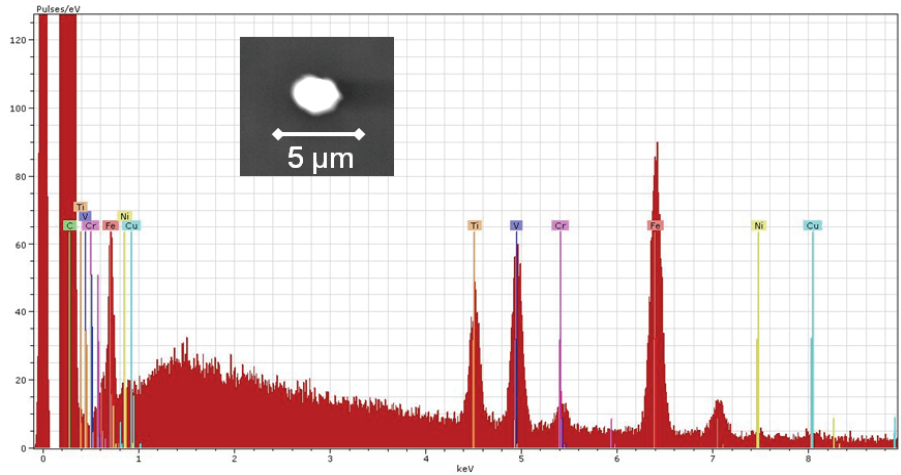

**Figure 6.** EDX spectrum of one of the particles found in a ZYA sample. Inset shows SEM image of the particle.

We estimate that a 2.5µm diameter particle of magnetite contributes ≈2.5·$10^{-9}$ emu to the overall magnetisation. Therefore, the observed ferromagnetic signal (1.5·$10^{-5}$ emu) for this particular ZYA sample (3×3×0.26 mm) implies that the sample contains ~6,000 magnetite particles which, if uniformly distributed, should be spaced by ~100 µm in the *ab* plane. This is in agreement with our SEM observations. This allows us to conclude that the visualized magnetic particles can indeed account for the whole ferromagnetic signal for this sample.

BS and EDX analysis of the other HOPG samples showing ferromagnetism produced similar results, with some samples containing predominantly Fe and others both Fe and Ti, as in the example above. A clear correlation has been found between the value of $M_S$ for a particular sample and the average separation of the magnetic particles detected by BS/EDX – see Fig. 7. No magnetic particles could be found in SPI samples and, accordingly, they did not show any ferromagnetic signal.

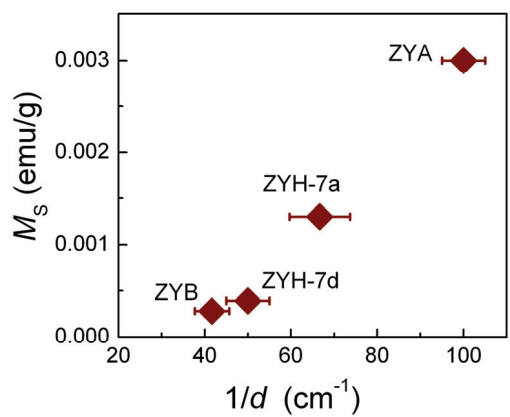

**Figure 7.** Saturation magnetisation $M_s$ as a function of the inverse of the average separation between the detected particles *d* as determined from the backscattering images (see text).

On the basis of the above analysis, we conclude that ferromagnetism in our ZYA, ZYB and ZYH HOPG samples is not intrinsic but due to contamination with micron-sized particles of probably either magnetite or titanomagnetite. As these particles were usually detected at submicron distances below the sample surface (see above), they should have been introduced during high-temperature crystal growth. We note that ZYA, ZYB or ZYH grades of HOPG are most commonly used for studies of magnetism in graphite (e.g., ZYA-grade crystals were specified in refs. [1, 4-7]) and, therefore, the contamination could be the reason for the reported ferromagnetism.

Finally, we would like to comment why magnetic particles such as those observed in the BS mode could have been overlooked by commonly used elemental analysis techniques, such as XRFS and PIXE [1-8]. Assuming that all particles found in our samples are magnetite and of approximately the same size, 2-3 µm, we estimate that the total number of Fe and Ti atoms in our samples ranges from 1 to 6 ppm. In the case of XRFS, 5 ppm of Fe is close to its typical detection limit and these concentrations might remain unnoticed. In the case of PIXE, its resolution is better than 1ppm. PIXE was used in e.g. refs. [1, 5] for ZYA graphite, where no contamination was reported but the saturation magnetisation was $\approx (1\text{-}2) \cdot 10^{-3}$ emu/g, similar to our measurements. The absence of detectable concentrations of magnetic impurities has been used as an argument that the ferromagnetic signals could not be due to contamination. Also, it was usually assumed that any remnant magnetic impurities were distributed homogeneously, rather than as macroscopic particles, and therefore would give rise to paramagnetism rather than a ferromagnetic signal, which was used as an extra argument against magnetic impurities.

It is clear that the latter assumption is incorrect, at least for the case of ZY grade graphite. Furthermore, let us note that the difference of several times for the limit put by PIXE and the amount measured by SQUID magnetometry is not massive. In our opinion, this difference can be explained by the fact that PIXE tends to underestimate the concentration of magnetic impurities if they are concentrated into relatively large particles. Indeed, PIXE probes only a thin surface layer (~1µm for 200 keV protons) which is thinner than the diameter of the observed magnetic inclusions. One can estimate that for round-shape inclusions with diameters of ~3µm, there should be a decrease by a factor of 3 in the PIXE signal with respect to the real concentration. Even more importantly, inclusions near the surface of HOPG provide weak mechanical points and are likely to be removed during cleavage when a fresh surface is prepared before PIXE analysis. Therefore, we believe that a micron-thick layer near the HOPG surface is unlikely to be representative of the whole sample. In contrast, magnetisation measurements probe average over the bulk of the samples, which can explain the observed several times discrepancy.

In conclusion, we have found that most commonly used crystals of highly-oriented pyrolytic graphite contain micron-size magnetic inclusions. The values of magnetization contributed by the detected inclusions are in agreement with the overall ferromagnetic moment measured for the corresponding crystals, indicating that the observed ferromagnetism has contamination origin. As HOPG from the same manufacturers and of the same grade have been used in many other studies of graphite's magnetic properties, magnetic inclusions are likely to be the reason for the often reported ferromagnetism in pristine graphite.

*Acknowledgements.* This work was supported by the UK Engineering and Physical Sciences Research Council. The authors are grateful to Michael Faulkner for his help with EDX and back-scattering SEM measurements.

*Corresponding author; e-mail: irina.grigorieva@manchester.ac.uk